\documentclass{aastex631}

\usepackage{amsmath,amssymb, amsthm,amstext}
\usepackage{graphicx}
\usepackage{color}
\usepackage{array, enumerate}
\usepackage{bm}
\usepackage{multirow}
\usepackage{braket}
\usepackage{txfonts}
\usepackage{booktabs}
\usepackage{afterpage}
\usepackage{subfigure}
\usepackage{rotating} 

\usepackage{ulem}
\usepackage{xcolor}

\newcommand{\old}[1]{{\color{red}{{}}}}
\newcommand{\reply}[1]{\textcolor{black}{ #1}}

\renewcommand{\cite}{\citep}

\begin{document}
\title{Lorentz Invariance Violation from Gamma-Ray Bursts\footnote{ H.~Song \& B.-Q.~Ma,  \href{https://doi.org/10.3847/1538-4357/adb8d4}{Astrophys.~J. 983 (2025) 9. }}}
\author[0009-0000-2543-8568]{Hanlin Song}
\affiliation{School of Physics, Peking University, Beijing 100871,China}
\author[0000-0002-5660-1070]{Bo-Qiang Ma}
\affiliation{School of Physics, Peking University, Beijing 100871,China}
\affiliation{Center for High Energy Physics, Peking University, Beijing 100871, China}
\affiliation{School of Physics, Zhengzhou University, Zhengzhou 450001, China}
\correspondingauthor{Bo-Qiang Ma}
\email{mabq@pku.edu.cn}

\begin{abstract}
Lorentz invariance violation (LV) is examined through the time delay between high-energy and low-energy photons in gamma-ray bursts (GRBs). Previous studies determined the LV energy scale as $E_{\rm LV} \simeq 3.60 \times 10^{17}$~GeV using Fermi Gamma-ray Space Telescope (FGST) data. This study updates the time-delay model and reaffirms these findings with new observations. High-energy photons from GRBs at GeV and TeV bands are analyzed, including the 99.3 GeV photon from GRB 221009A (FGST), the 1.07 TeV photon from GRB 190114C (MAGIC), and the 12.2 TeV photon from GRB 221009A (LHAASO). Our analysis, in conjunction with previous data, consistently shows that high-energy photons are emitted earlier than low-energy photons at the source. By evaluating 17 high-energy photons from 10 GRBs observed by FGST, MAGIC, and LHAASO, we estimate the LV energy scale to be $E_{\rm LV} \simeq 3.00 \times 10^{17}$ GeV. The null hypothesis of dispersion-free vacuum $E=pc$ (or, equivalently, the constant light-speed $v_{\gamma}=c$) is rejected at a significance level of 3.1$\sigma$ or higher.
\end{abstract} 

\section{Introduction} 
Lorentz invariance has been a foundational principle in modern physics since the inception of relativity.
 However, there exist theories that allow for Lorentz invariance violation (LV; for recent reviews, see, e.g. \cite{He:2022gyk,AlvesBatista:2023wqm}),
such as string theory \citep{Amelino-Camelia:1996bln, Amelino-Camelia:1997ieq, Ellis:1999rz, Ellis:1999uh, Ellis:2008gg, Li:2009tt, Li:2021gah, Li:2021eza}, 
loop quantum gravity \citep{Gambini:1998it, Alfaro:1999wd,Li:2022szn}, and doubly special relativity \citep{Amelino-Camelia:2002cqb, Amelino-Camelia:2002uql, Amelino-Camelia:2000stu}. 
These theories predict that LV occurs around the Planck scale 
$E_{\rm P} \equiv \sqrt{\hbar c^5/G} \simeq 1.22 \times 10^{19}$ GeV. 
For instance, when the energy of a photon is significantly less than the Planck scale, the modified dispersion relation needs to be adjusted in a model-independent manner 
as 
\citep{Xiao:2009xe, He:2022gyk}
\begin{equation}
E^2 \simeq p^2c^2\left[1-s_n\left(\frac{pc}{E_{\mathrm{LV},n}}\right)^n\right],
\end{equation}
where $p$ represents  the momentum of the photon, $E_{{\rm{LV}},n}$ denotes  the $n$th-order energy scale of Lorentz invariance violation or light-speed variation (where either case can be abbreviated as ``LV''),  and $s_n \equiv \pm 1$ indicates whether high-energy photons travel faster ($s_n = -1$) or slower ($s_n = +1$) than low-energy photons. By applying the relation $v = \partial E / \partial p$, the group velocity of the photon can be expressed as: 
\begin{equation}
v \simeq c\left[1-s_n\frac{n+1}{2}\left(\frac{pc}{E_{\mathrm{LV},n}}\right)^n\right],
\end{equation}
which implies a tiny light-speed variation. 
However, due to the significant suppression from $E_{{\rm{LV}},n}$
near the Planck scale, observing LV effects directly through experiments on Earth is challenging.
Amelino-Camelia {\it et al.}~\citep{Amelino-Camelia:1996bln, Amelino-Camelia:1997ieq}  
first proposed using the time delay between high-energy photons and low-energy photons from gamma-ray bursts (GRBs) to investigate LV. The vast cosmological distances traveled by these photons can magnify even slight velocity differences into observable time delays. 
Accounting for the expansion of the Universe, the observed time delay on Earth can be divided into two components 
\citep{Ellis:2005sjy, Shao:2009bv, Zhang:2014wpb, Xu:2016zxi, Xu:2016zsa, Huang:2019etr, Zhu:2021pml, Zhu:2021wtw, Zhu:2022usw}
\begin{equation}
\label{obsdelay}
\Delta t_{\mathrm{obs}}=\Delta t_{\mathrm{LV}}+(1+z)\Delta t_{\mathrm{in}},
\end{equation}
where $\Delta t_{\mathrm{obs}}$ represents the observed time delay by scientific observatories, $\Delta t_{\mathrm{LV}}$ denotes the time delay caused by LV, $\Delta t_{\mathrm{in}}$ is the intrinsic time delay at the source frame, and $z$ 
signifies the redshift of the corresponding GRB.
The time delay resulting from LV can be expressed as
\citep{Jacob:2008bw, Zhu:2022blp}
\begin{equation}
\label{lorentzdelay}
\Delta t_{\mathrm{LV}}=s_{n}\frac{1+n}{2H_{0}}\frac{E_{\mathrm{high,o}}^{n}-E_{{\rm low,o}}^{n}}{E_{\mathrm{LV},n}^{n}}\int_{0}^{z}\frac{(1+z')^{n}\mathrm{d}z'}{\sqrt{\Omega_{m}(1+z')^{3}+\Omega_{\Lambda}}},
\end{equation}
where $E_{\mathrm{high,o}}^{n}$ and $E_{\rm low,o}^{n}$ correspond to the energies of high- and low-energy photons and $H_0$, $\Omega_{m}$, and $\Omega_{\Lambda}$ represent  the Hubble constant, matter density parameter, and dark energy density parameter of the $\Lambda$CDM model, respectively.  

Previous studies~\citep{Xu:2016zxi, Xu:2016zsa}  analyzed eight GRBs with 14 high-energy photon events ($>$10 GeV) observed by the Fermi Gamma-ray Space Telescope (FGST). The results suggested that $n=1$, $s_n = + 1$, $E_{\rm LV,1} \simeq 3.60 \times 10^{17}$ GeV, with the conclusion supported by more detailed analyses \citep{Liu:2018qrg, Xu:2018ien, Chen:2019avc, Li:2020uef,  Zhu:2021pml, Zhu:2021wtw}. 
Additionally, a negative intrinsic time delay was indicated from fitting data, implying that high-energy photons are emitted before low-energy photons \citep{Chen:2019avc, Zhu:2021pml, Zhu:2021wtw}. 
However, previous works assumed a constant intrinsic time delay for all high-energy photon events. 
Recently, an increasing number of high-energy photons from GRBs with known redshifts have been observed across the GeV to TeV bands, 
for instance,
the 99.3 GeV photon of GRB 221009A observed by FGST 
\citep{Lesage:2023vvj, Zhu:2022usw}, the 1.07 TeV photon of GRB 190114C observed by the Major Atmospheric Gamma Imaging Cherenkov (MAGIC) telescope \citep{MAGIC:2019lau, MAGIC:2020egb}, and the 12.2 TeV photon of GRB 221009A observed by the Large High Altitude Air-shower Observatory (LHAASO) \citep{LHAASO:2023lkv}.

Given the wide energy range of these high-energy photons from 33.58 GeV to 14.04 TeV at the source frame, it is plausible that these photons are emitted at different times.

Therefore, in this study, we refine the intrinsic time-delay model~\cite{plb138951} to consider the influence of a broad range of source energies. Our analysis aligns with previous work \citep{Xu:2016zxi, Xu:2016zsa} 
after examining the same GRB data
(labeled I in Table~\ref{HighEnergyPhotons}). 
Furthermore, we evaluate previous GRB data alongside data from three new GRBs
(labeled II in Table~\ref{HighEnergyPhotons}) observed by FGST, MAGIC, and LHAASO, respectively. 
All scenarios yield consistent results and support the same physical scenario. Finally, by analyzing all 17 high-energy photons from 33.58 GeV to 14.04 TeV at the source frame of 10 GRBs observed by FGST, MAGIC, and LHAASO, our results suggest that high-energy photons are emitted before low-energy photons, indicating an LV energy scale at $E_{\rm LV} \simeq 3.00 \times 10^{17}$ GeV.

\section{Observed GRB data}
\label{GRBData}
In this study, we analyze the high-energy photons ($>$10 GeV) observed in GRB data from FGST, MAGIC, and LHAASO, as presented in Table~\ref{HighEnergyPhotons}.
In this table, label I represents 14 high-energy photons from eight GRBs analyzed in previous studies \citep{Xu:2016zxi, Xu:2016zsa}, with energies ranging from above 10 GeV to below 80 GeV. Label II denotes three high-energy photons newly observed by FGST, MAGIC, and LHAASO, with energies ranging from 99.3 GeV (new FGST data from GRB 221009A) to approximately 12.2 TeV (the highest energy photon in LHAASO data from GRB 221009A).

In fact, the LHAASO highest energy photon was initially attributed as 18~TeV upon discovery~\cite{LHAASO}
but refined analysis later on shifted the energy 
value to  \(E = 12.2^{+3.5}_{-2.4}\) TeV~\cite{LHAASO:2023lkv}.
The redshift 
$z$, observed energy of high-energy photons in the observer frame 
$E_{\rm high,o}$, and arrival time of high-energy photons 
$T_{\rm high,o}$ are acquired from various observatories.

\begin{figure}[ht!]
\centering
	\includegraphics[width=0.97\textwidth]{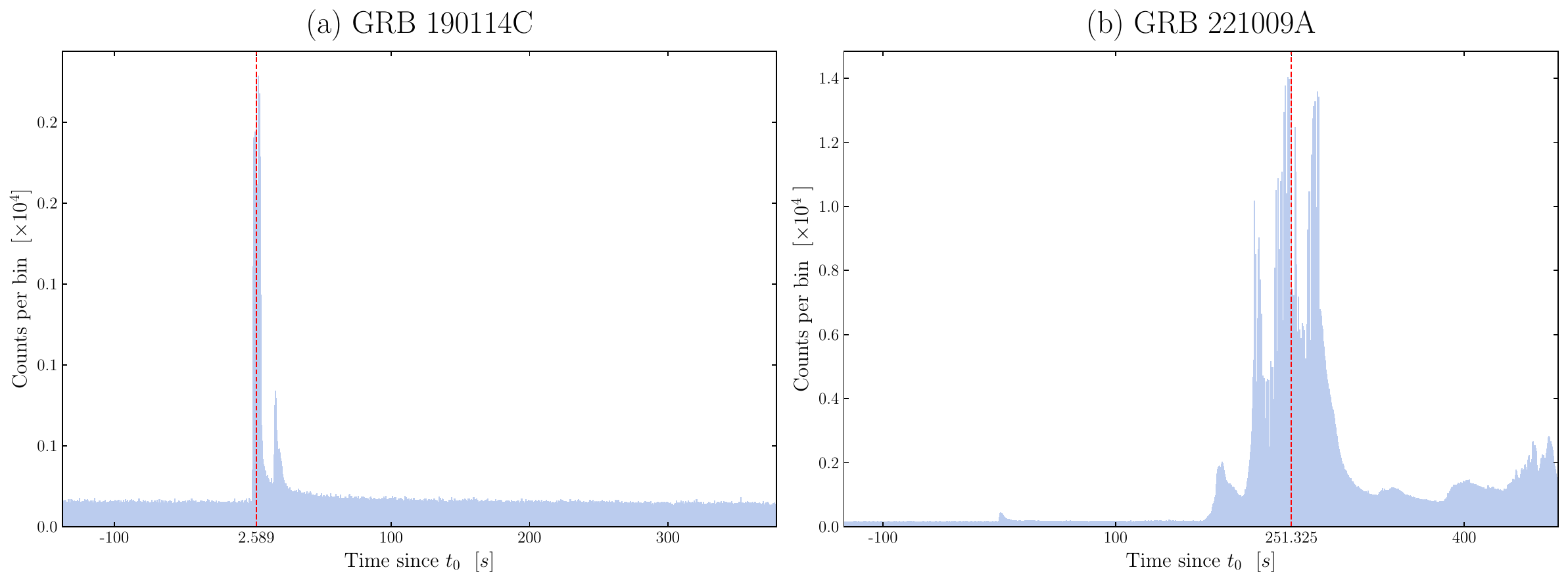} 
	\caption{Light curve of GBM data with the two brightest trigger detectors combined for GRB 190114C (GBM NaI-n7 and NaI-n8, 8-260 keV) and GRB 221009A (GBM NaI-n7 and NaI-n8, 8-260 keV). These histograms are binned in 64 ms, consistent with \citep{Xu:2016zxi, Xu:2016zsa}. The red dashed line in each subfigure indicates the first significant peak of low-energy photons. Here, we define time 0 as the GBM trigger time  $t_0$  for GRB 190114C and GRB 221009A, respectively.} 
	\label{GBM} 
\end{figure}

Following the methodology of previous works \citep{Shao:2009bv, Zhang:2014wpb, Xu:2016zxi, Xu:2016zsa}, we consider low-energy photons ranging from 8 keV to 260 keV observed by the Gamma-ray Burst Monitor (GBM) \citep{Meegan:2009qu} as the low-energy photon reference for comparison with high-energy photons in our calculations.
The arrival time of low-energy photons $T_{\rm low,o}$ 
is determined as the first main significant peak \citep{Xu:2016zxi, Xu:2016zsa, Liu:2018qrg}. 
We use the same  $T_{\rm low,o}$ values
for the previous eight GRBs as in previous studies \citep{Xu:2016zxi, Xu:2016zsa}. 
For the two new  GRBs, the $T_{\rm low,o}$ values are obtained as depicted in Fig.~\ref{GBM}. 
To account for uncertainty, we consider a positional uncertainty of  $\pm 5$~s 
in the observer frame for the first main significant peak. 
Different choices of the uncertainty, such as of $\pm 1$~s or $\pm 10$~s, only alter the results slightly.
The observed time delay in the observer frame is then calculated as
\begin{equation}
    \Delta t_{\mathrm{obs}} = t_{\rm high,o} - t_{\rm low,o}.
\end{equation}

Consistent with previous works \citep{Xu:2016zxi, Xu:2016zsa}, 
we take $n=1$ and $s_n=+1$, Eq.~\ref{obsdelay} is then
reformulated as
\begin{equation}
\label{sourcedelay}
\frac{\Delta t_{\mathrm{obs}}}{1+z} = \Delta t_{\mathrm{in}} + a_{\rm LV}K_1,
\end{equation}
where $a_{\rm LV} = 1/E_{\rm LV}$ and $K_1$ is
\begin{equation}
K_1=\frac{1}{H_0}\frac{E_{\rm high,o}}{1+z}\int_0^z\frac{(1+z')\mathrm{d}z'}{\sqrt{\Omega_\mathrm{m}(1+z')^3+\Omega_\Lambda}}.
\end{equation}
Here, we take $E_{\rm low,o}$ (8-260~keV) as 0 
due to its significantly lower energy compared to 
$E_{\rm high,o}$ (11.9-12200~GeV). 
The energy resolution varies among different observatories, with FGST having $\pm 10\%$ resolution for high-energy photons from 10 GeV to hundreds of GeV \citep{Fermi-LAT:2009pgs}, MAGIC with $\pm 15\%$ 
resolution for observed TeV photons \citep{MAGIC:2019lau}, and LHAASO with a 
$\pm 20\%$ resolution for high-energy photons from 10 TeV to 16 PeV \citep{LHAASO:2023lkv}. 
For cosmological parameters, we adopt the Planck 2018 results for
$H_0 = 67.36 \pm 0.54 \ \rm{km \ s^{-1} \ Mpc^{-1}}$ , $\Omega_{m} = 0.3153 \pm 0.0073$, and $\Omega_{\Lambda} = 0.6847 \pm 0.0073$ \citep{Planck:2018vyg}.
The uncertainties and values of $\Delta t_{\mathrm{obs}} / {(1+z)}$,
the high-energy photon values in the source frame  $E_{\rm high,s}$, and $K_1$ in Table~\ref{HighEnergyPhotons} are then determined.

\section{Intrinsic time-delay models and parameter estimation methods}

\begin{figure}[htbp] 
    \centering
	\includegraphics[width=0.75\textwidth]{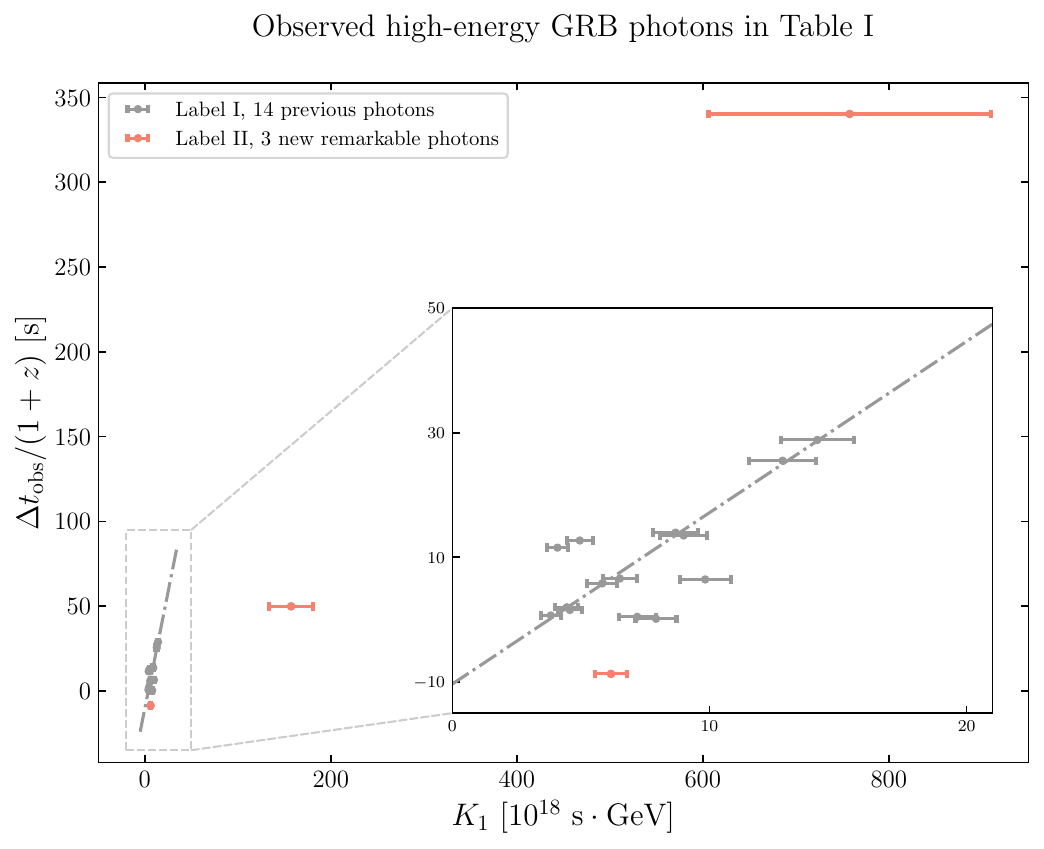} 
	\caption{\reply{The 17 observed high-energy photons in Table~I. The gray dots denote the 14 photons in lable I, which were analyzed in~\cite{Xu:2016zxi,Xu:2016zsa,plb138951}. The red dots denote the three new remarkable photons in  lable II, which are the 99.3 GeV photons observed by FGST \cite{Lesage:2023vvj}, the 1.07 TeV photon observed by MAGIC \cite{MAGIC:2019lau, MAGIC:2020egb}, and the 12.2 TeV photon observed by LHAASO \cite{LHAASO:2023lkv} from bottom to top. The events in the zoomed-in subfigure (the ticks of the x- and y-axes are rescaled for clarity) denote the GeV-band photons, which can be fitted with a linear relationship, as pointed out in~\cite{Xu:2016zxi,Xu:2016zsa,plb138951}.}} 
	\label{Observed_GRB_photons} 
\end{figure}

In previous works \citep{Shao:2009bv, Zhang:2014wpb, Xu:2016zxi, Xu:2016zsa, Huang:2019etr, Zhu:2021pml, Zhu:2021wtw, Zhu:2022usw}, 
the intrinsic time delay was treated as a constant term common to all GeV-band photons. \reply{
As shown in the zoomed-in subfigure (the ticks of the x- and y-axes are rescaled for clarity) of Fig.~\ref{Observed_GRB_photons}, an apparent linear relationship for GeV-band photons can be fingered out. The slope of the dotted-dashed line denotes the Lorentz invariance violation parameters $a_{\rm LV}$, and the interception denotes the common intrinsic time delay  $\Delta t_{\rm in, c}$. }

\reply{
However, the new observed TeV photons of GRB 190114C from MAGIC \cite{MAGIC:2019lau, MAGIC:2020egb} and GRB 221009A from LHAASO \cite{LHAASO:2023lkv} show an apparent derivation from the linear relationship of the GeV band photons.  This motivates us to adopt a more detailed intrinsic time-delay model. Previous work in~\cite{plb138951} introduced an energy-dependent intrinsic time-delay model for GeV band photons, which revealed a similar physical scenario for intrinsic properties and a consistence LV energy scale $E_{\rm LV} \sim 3 \times 10^{17}$ GeV, compared to the previous model. Therefore, in this study, we explore the applicability of the energy-dependent intrinsic time-delay model for both GeV-band and TeV-band photons.}
The intrinsic time-delay term is now expressed as
\begin{equation}
    \Delta t_{\rm in} = \Delta t_{\rm in, c} + \alpha E_{\rm high,s},
\end{equation}
where $\alpha$ represents the coefficient and 
the contribution from low-energy photons is disregarded due to their considerably lower energy values.
Subsequently, Eq.~\ref{sourcedelay} is reformulated as
\begin{equation}
\label{model}
    \frac{\Delta t_{\mathrm{obs}}}{1+z} = a_{\rm LV}K_1 + \alpha E_{\rm high,s} + \Delta t_{\rm in, c}.
\end{equation}

We then utilize a general linear model in which the true value of 
$y_{\rm t}$ depends on the true values of 
$x_{{\rm t}1}$ and $x_{{\rm t}2}$ as
\begin{equation}
    y_{\rm t} = \beta_1 x_{{\rm t}1} + \beta_2 x_{{\rm t}1} + \beta_0,
\end{equation}
where $\beta_0$, $\beta_1$, and $\beta_2$ are the coefficients 
to be determined by taking into account measurement errors for physical variables. 
Due to the presence of noise, there are inherent measurement errors associated with physical variables,
\begin{equation}
    \begin{cases}
        y_{\rm m} = y_{\rm t} + y_{\rm e}, \\
        x_{\rm m1} = x_{\rm t1} + x_{\rm e1},\\
        x_{\rm m2} = x_{\rm t2} + x_{\rm e2},
    \end{cases}
\end{equation}
where ${y_{\rm m}, x_{\rm m1}, x_{\rm m2}}$ and ${y_{\rm e}, x_{\rm e1}, x_{\rm e2}}$ 
represent the measurement values and measurement errors of the variables, respectively. It is assumed that these measurement error variables follow a Gaussian distribution. 
Given that the corresponding coefficients ${\beta_0, \beta_1, \beta_2}$ and $q$ sets of ${x_{\rm m1}, x_{\rm m2}}$ are known, the likelihood function for measuring $q$ times of $y_{\rm m}$ is
\begin{equation}
\begin{aligned}
     p\left(\{y_{\rm m}\} \mid \beta_0, \beta_1, \beta_2, \{x_1,x_2\}\right) =  
     \prod_{j=1}^{q} \frac{1}{\sqrt{2\pi\left(\beta_1^2\sigma_{x_{\rm m1},j}^2 + \beta_2^2\sigma_{x_{\rm m2},j}^2 + \sigma_{y_{{\rm m},j}}^2 \right)}}   
     \exp{\left(-
 \frac{\left(y_{{\rm m},j} - \beta_1x_{{\rm m1,j}} - \beta_2x_{{\rm m2,j}} - \beta_0\right)^2}{2\left(\beta_1^2\sigma_{x_{\rm m1},j}^2 + \beta_2^2\sigma_{x_{\rm m2},j}^2 + \sigma_{y_{{\rm m},j}}^2 \right)}\right)}.
\end{aligned}
\end{equation}
Then, according to the Bayesian theorem, the posterior is
\begin{equation}
\begin{aligned}
     p\left({\beta_0,\beta_1, \beta_2 \mid \{y_{\rm m}\} , \{x_{\rm 1},x_{\rm 2}\}} \right) \propto  \
     p\left({\{y_{\rm m}\} \mid  \{x_{\rm 1},x_{\rm 2}\}, \beta_0,\beta_1, \beta_2} \right) p\left(\beta_0\right)p\left(\beta_1\right)p\left(\beta_2\right).
\end{aligned}
\end{equation}

    In the context of the physical model presented in this study, $y_{\rm t}$ corresponds to ${\Delta t_{\mathrm{obs}}}/{(1+z)}$, while $x_{\rm t1}$ and $x_{\rm t2}$ are related to $K_1$ and $E_{\rm high,s}$ in Eq.~9, respectively.
Meanwhile, 
    $\Delta t_{\rm in,c}$ represents the common intrinsic time-delay emission term for high-energy photons in Eq.~9. We assume it follows a Gaussian distribution, allowing the emission to occur over a finite time interval (denoted by $\sigma$), rather than as an instantaneous peak in the form of a $\delta$ function:
    
\begin{equation}
    p\left(\Delta t_{\rm in,c} \right) \sim \mathcal{N} \left(\mu, \sigma^2\right).
\end{equation}

Then there are four parameters that need to be determined: the LV parameter $a_{\rm LV}$, the energy-dependent coefficient $\alpha$, the mean value of the common intrinsic time delay $\mu$, and the variance $\sigma$. We assume that all these parameters follow flat distributions:
\begin{align}
    \begin{cases}
        p\left(a_{\rm LV}\right) \sim U\left[0, 30 \right] \times 10^{-18} \ \rm{GeV^{-1}}, \\
        p(\alpha) \sim U \left[-30, 30 \right] \ {\rm s \cdot GeV^{-1}}, \\
        p\left(\mu\right) \sim U\left[-30, 30 \right] \ {\rm s}, \\
        p\left(\sigma\right) \sim U\left[0, 30 \right] \ {\rm s}. \\
    \end{cases}
    \label{priors}
\end{align}
After marginalizing over the common time-delay term $\Delta t_{\rm in,c}$, the posterior finally becomes:
\begin{equation}
\begin{aligned}
     p \propto\prod_{j=1}^{q} \frac{1}{\sqrt{2\pi\left(a_{\rm LV}^2\sigma_{K_{1,j}}^2 + \alpha^2\sigma_{E_{{\rm high,s},j}}^2 + \sigma^2 + \sigma_{y_{j}}^2 \right)}}  
     \exp{\left(-
 \frac{\left(\frac{\Delta t_{\mathrm{obs},j}}{1+z_{j}} - a_{\rm LV}{K_{1,j}} - \alpha E_{{\rm high,s},j} - \mu\right)^2}{2\left(a_{\rm LV}^2\sigma_{K_{1,j}}^2 + \alpha^2\sigma_{E_{{\rm high,s},j}}^2 + \sigma^2 + \sigma_{y_{j}}^2 \right)}\right)}p\left(a_{\rm LV}\right)p(\alpha)p\left(\mu\right)p\left(\sigma\right).
\end{aligned}
\end{equation}

Subsequently, by utilizing the measured data of GRBs from Table~\ref{HighEnergyPhotons}, we can determine the posterior distributions of these four parameters. 
We perform the parameter estimation procedure based on the Bayes theorem, a method widely used in various areas of physics, such as gravitational-wave data analysis \cite{ LIGOScientific:2016aoc, LIGOScientific:2018jsj, KAGRA:2021duu} and testing the Lorentz invariance violation with GRBs \cite{Pan:2020zbl, Vardanyan:2022ujc}. 
We employ 
the \texttt{bilby} \citep{Ashton:2018jfp, Romero-Shaw:2020owr} package for our 
computations.

We conducted a Monte Carlo simulation using simulated data that includes 10 GRBs with the same redshifts as those in our analyzed realistic dataset. Assuming that all GRBs share similar intrinsic properties, we utilized the intrinsic spectra derived from the observation of GRB 221009A as reported by LHASSO-WCDA \cite{LHAASO:2023kyg}. For each GRB, we randomly generated 1000 high-energy photons in the source frame, with energies $E_{\rm high,s}$ ranging from 2 to 20,000 GeV.
Subsequently, we created four simulated datasets across different scenarios by injecting parameters obtained from real data analysis, incorporating the observed time delay for a given photon with energy $E_{\rm high,s}$. The time delay is calculated using the following equation:
\begin{equation}
    \label{model}
    \frac{\Delta t_{\mathrm{obs}}}{1+z} = a_{\rm LV}K_1 + \alpha E_{\rm high,s} + \Delta t_{\rm in, c}, \  p(\Delta t_{\rm in, c}) \sim \mathcal{N}(\mu, \sigma^2).
\end{equation}
Our results demonstrate that we can effectively reproduce the injected parameters with high efficiency. This outcome illustrates the robustness of our approach in capturing a broad range of emission dynamics, thereby enhancing our sensitivity to potential signals of Lorentz invariance violation. The ability to accurately recover these parameters reinforces the reliability of our simulation methodology and supports the validity of our findings in the context of LV searches.

\section{Results and discussion}
\begin{figure}[htbp] 
    \centering
	\includegraphics[width=0.95\textwidth]{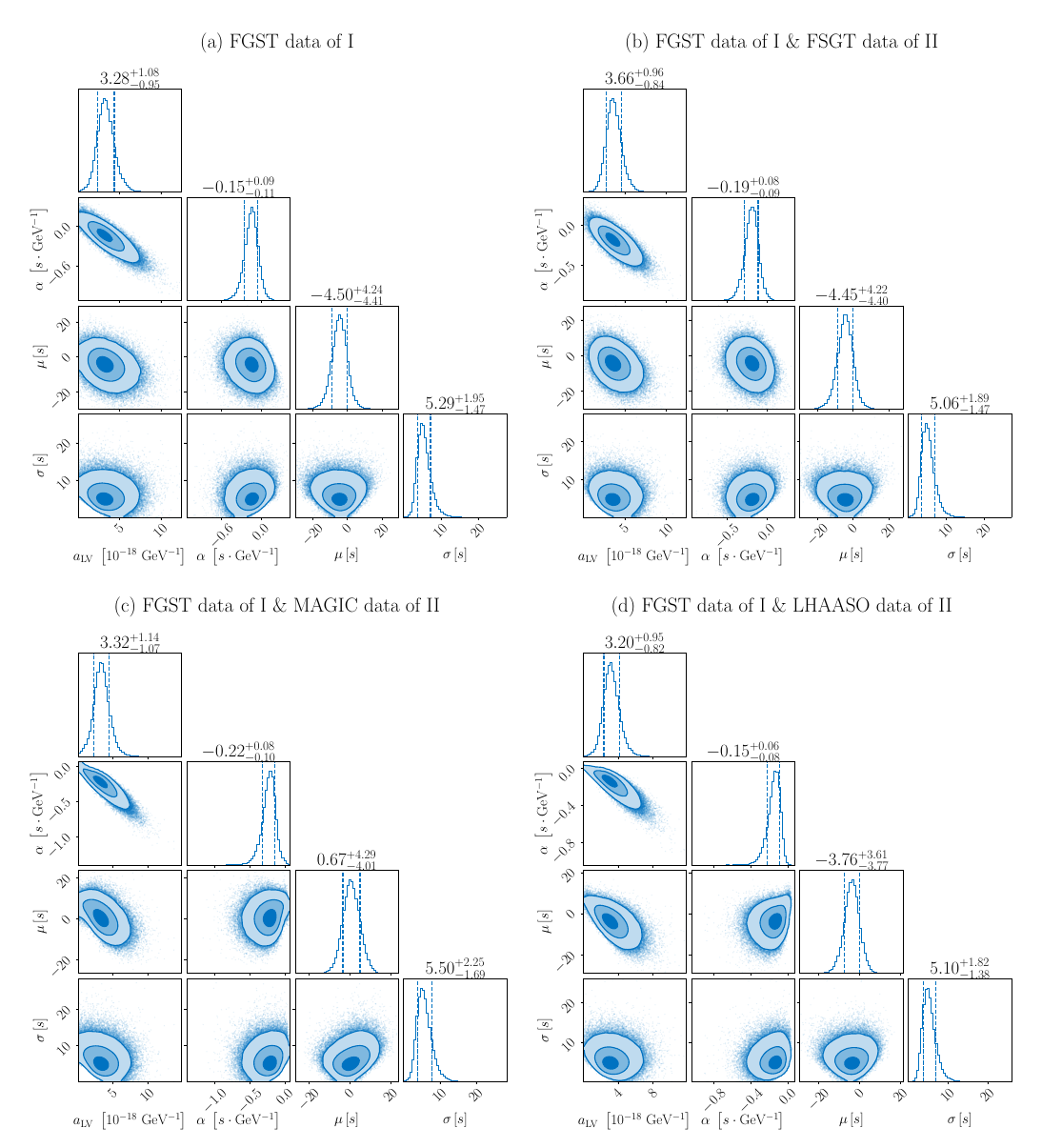} 
	\caption{
Posterior distributions for four cases are presented. Panel (a) displays the results for the GRB data in label I. Panel (b) illustrates the results for the combined data of GRBs in label I and the new FGST data from label II. Panel (c) showcases the results for the combined data of GRBs in label I and the new MAGIC data from label II. Panel (d) demonstrates the results for the combined data of GRBs in label I and the new LHAASO data from label II. The 2D contours represent different confidence levels, denoting the 1 $\sigma$, 2 $\sigma$, and 3 $\sigma$ regions, while the vertical lines indicate the  1 $\sigma$ region for the 1D marginalized posterior distribution.
 } 
    \label{PM} 
\end{figure}

As depicted in panel (a) of Fig.~\ref{PM}, we reanalyze the GRBs data from label I, which comprises the same raw data utilized in previous studies \citep{Xu:2016zxi,Xu:2016zsa}, but incorporating the new intrinsic time-delay model \cite{plb138951} with updated cosmological parameters and considering the uncertainty in the arrival time of low-energy photons.
The results indicate the LV parameter $a_{\rm LV} = 3.28^{+1.08}_{-0.95} \times 10^{-18} \ \rm{GeV^{-1}}$, the energy-dependent parameter $\alpha = -0.15^{+0.09}_{-0.11} \ {\rm s \cdot GeV^{-1}}$, the mean value of common intrinsic time delay $\mu = -4.50^{+4.24}_{-4.41}$~s, and the variance of the common intrinsic time delay  $\sigma = 5.29^{+1.95}_{-1.47}$~s. 
These results suggest a physical scenario where high-energy photons are emitted earlier than low-energy photons, with the LV energy scale estimated at $E_{\rm LV} = 3.05^{+ 1.24}_{-0.75} \times 10^{17}$ GeV. 
This outcome is consistent with previous works~\citep{Xu:2016zxi,Xu:2016zsa} in terms of both the physical scenario and the LV energy scale ($E_{\rm LV} = 3.60 \pm 0.26 \times 10^{17}$ ~GeV as reported therein).

Subsequently, we combine the GRB data from label I with each new GRB datum from label II individually. 
In panels (b), (c), and (d) of Fig.~\ref{PM}, all analyses yield consistent physical scenarios and LV energy scales
\begin{itemize}
\item 
In panel (b), we examine the 14 previous high-energy photons from FGST along with a new 99.3~GeV photon from GRB 221009A observed by FGST.
The results suggest the LV parameter $a_{\rm LV} = 3.66^{+0.96}_{-0.84} \times 10^{-18} \ \rm{GeV^{-1}}$, the energy-dependent parameter $\alpha = -0.19^{+0.08}_{-0.09} \ {\rm s \cdot GeV^{-1}}$, the mean value of common intrinsic time delay $\mu = -4.45^{+4.22}_{-4.40}$~s, and the variance of  the common intrinsic time delay  $\sigma = 5.06^{+1.89}_{-1.47}$~s. 
The corresponding LV energy scale is estimated at
$E_{\rm LV} = 2.74^{+ 0.82}_{-0.57} \times 10^{17}$ GeV. 
\item 
In panel (c), we analyze the 14 previous high-energy photons from FGST in addition to a new 1.07 TeV photon from GRB 190114C observed by MAGIC.
The results suggest the LV parameter $a_{\rm LV} = 3.32^{+1.14}_{-1.07} \times 10^{-18} \ \rm{GeV^{-1}}$, the energy-dependent parameter $\alpha = -0.22^{+0.08}_{-0.10} \ {\rm s \cdot GeV^{-1}}$, the mean value of common intrinsic time delay $\mu = 0.67^{+4.29}_{-4.01}$~s, and the variance of the common intrinsic time delay  $\sigma = 5.50^{+2.25}_{-1.69}$~s. 
The corresponding LV energy scale is 
estimated at 
$E_{\rm LV} = 3.01^{+ 1.44}_{-0.77} \times 10^{17}$ GeV.  
Although the mean value $\mu$ is slightly positive here, 
the negative value of $\alpha$ indicates the same physical scenario where high-energy photons are emitted earlier than low-energy photons. 
\item 
In panel (d), 
we analyze the 14 previous high-energy photons from FGST alongside a new 12.2~TeV photon from GRB 221009A observed by LHAASO. 
The results suggest the LV parameter $a_{\rm LV} = 3.20^{+0.95}_{-0.82} \times 10^{-18} \ \rm{GeV^{-1}}$, the energy-dependent parameter $\alpha = -0.15^{+0.06}_{-0.08} \ {\rm s \cdot GeV^{-1}}$, the mean value of common intrinsic time delay $\mu = -3.76^{+3.61}_{-3.77}$~s, and the variance of the common intrinsic time delay  $\sigma = 5.10^{+1.82}_{-1.38}$~s. The corresponding LV energy scale is estimated at $E_{\rm LV} = 3.13^{+ 1.08}_{-0.72} \times 10^{17}$ GeV. 
\end{itemize}
Consequently, all results support the physical scenario where high-energy photons are emitted earlier in the source frame, indicating an LV energy scale around $E_{\rm LV} \sim 3 \times 10^{17}$ GeV.

\begin{figure}[htbp] 
    \centering
	\includegraphics[width=0.60\textwidth]{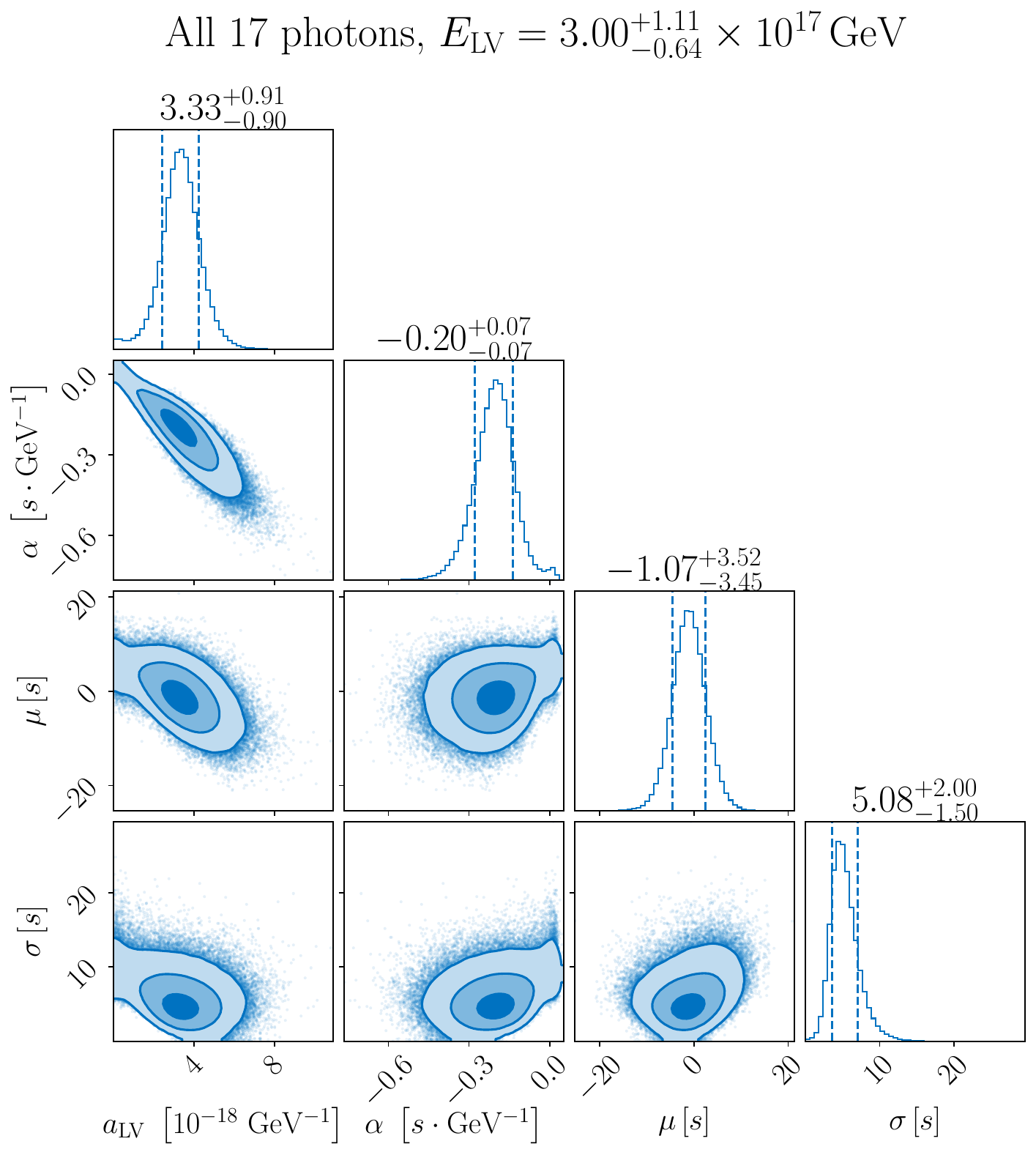} 
	\caption{
 The posterior distributions are shown for the combined analysis of all 17 high-energy photons ranging from 33.58 GeV to 14.04 TeV at the source frame of 10 GRBs observed by FGST, MAGIC, and LHAASO.
 } 
	\label{combian} 
\end{figure}

Lastly, we analyze all 17 high-energy photons from 33.58 GeV to 14.04 TeV at the source frame of 10 GRBs observed by FGST, MAGIC, and LHAASO, and the results are displayed in Fig.~\ref{combian}. 
Once again, we obtain consistent outcomes indicating 
the LV parameter $a_{\rm LV} = 3.33^{+0.91}_{-0.90} \times 10^{-18} \ \rm{GeV^{-1}}$, the energy-dependent parameter $\alpha = -0.20^{+0.07}_{-0.07} \ {\rm s \cdot GeV^{-1}}$, the mean value of common intrinsic time delay $\mu = -1.07^{+3.52}_{-3.45}$ ~, and the variance of the the common intrinsic time delay  $\sigma = 5.08^{+2.00}_{-1.50}$~s. 
These results support the physical scenario where high-energy photons are emitted earlier than low-energy photons, with the LV energy scale estimated at 
$E_{\rm LV} = 3.00^{+ 1.11}_{-0.64} \times 10^{17}$ GeV. 

In our calculations, we have considered both GRBs from a sample of eight events and short GRBs, specifically GRB 090510 and GRB 140619B. While, in principle, these two types of GRBs should be analyzed separately due to their distinct characteristics, our results indicate only minor differences when comparing the analysis based solely on the eight long GRBs to that which includes all 10 GRBs (both long and short). This suggests that our assumption of similar intrinsic properties among the 10 GRBs is a reasonable first approximation for the purposes of this study.
In principle, we cannot apply this model to photons from an individual GRB due to the degeneracy between the parameters \(a_{\rm LV}\) and \(\alpha\), as highlighted in \cite{plb138951}. Analyzing a single GRB event necessitates the incorporation of prior information from multiple GRBs to effectively break this degeneracy. To address this, we first analyze a dataset comprising 8 + 3 photons from nine GRBs (excluding the six photons from GRB 090902B) and obtain results for \(E_{\rm LV}\) and \(\alpha\) that are consistent with those derived from fitting the complete 17-photon dataset. Subsequently, we use the obtained value of \(\alpha\) as a calibration parameter to analyze the six photons from GRB 090902B, which allows us to reconstruct \(E_{\rm LV} \sim \mathcal{O}(10^{17})\) GeV consistently.

Our prediction of a preburst stage for the emission of many high-energy photons before the prompt burst of low-energy photons is further supported by the data depicted in Fig.~1 of the LHAASO
observation of multi-TeV photons from GRB 221009A~\cite{LHAASO:2023lkv}. Specifically, multi-TeV photons with energies
exceeding 3 TeV were observed within a time frame of -70 to 1400 s relative to the trigger time~\cite{Liu:2024qbt}.
This data set unequivocally demonstrates the detection of a number of multi-TeV photons preceding the
emergence of keV-MeV photons, multi-GeV photons, and multi-hundred-GeV photons~\cite{Liu:2024qbt}. These observations provide direct evidence in support of the idea that a significant emission of multi-TeV photons
occurs prior to the onset of lower-energy photons during the prompt burst phase at the GRB source,
even in the absence of Lorentz invariance violation. This feature can elucidate why our results about the Lorentz violation scale differ from those more stringent constraints in the literature~\cite{FermiGBMLAT:2009nfe,Xiao:2009xe,Vasileiou:2013vra,LHAASO:2024lub}.

\begin{figure}[htbp] 
    \centering
	\includegraphics[width=0.65\textwidth]{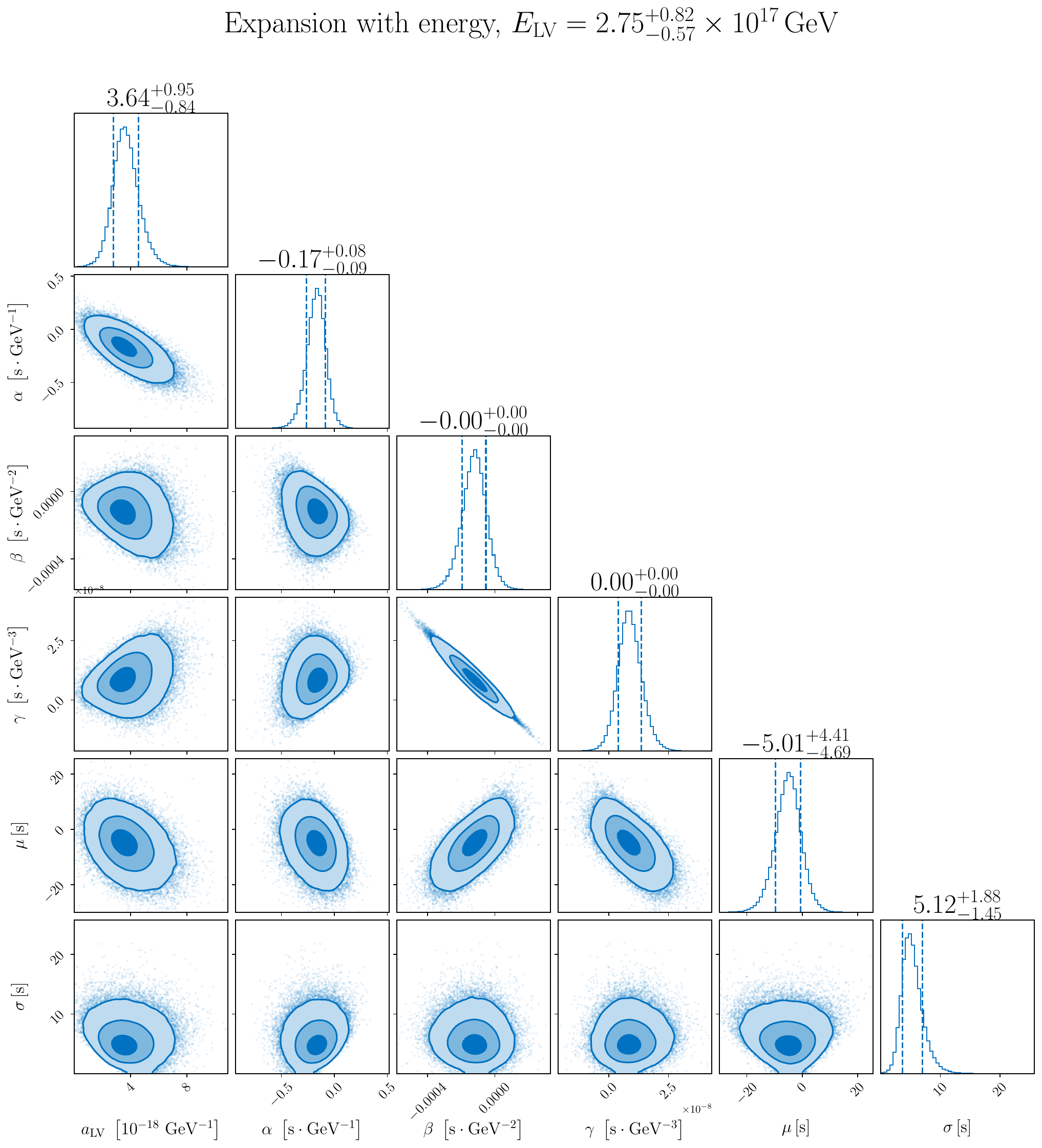} 
	\caption{
The posterior distributions illustrating the combined analysis of all 17 high-energy photons from 10 GRBs observed by FGST, MAGIC, and LHAASO. The analysis incorporates the general expression of intrinsic photon emission time as given in Eq.~\ref{generalDelta}, showcasing the convergence of both coefficients 
$\beta$ and $\gamma$ towards 0 during data fitting. This convergence suggests the validity of a linear relationship between energy and intrinsic photon emission time. The priors for $a_{\rm LV}$, $\alpha$, $\mu$ and $\sigma$ are the same as in Eq.~\ref{priors}, and $\beta$ and $\gamma$ are sampled from uniform distribution $U \left[-30, 30 \right] \ {\rm s \cdot GeV^{-2}}$ and $U \left[-30, 30 \right] \ {\rm s \cdot GeV^{-3}}$, respectively.}
	\label{energy_taylor} 
\end{figure}

Our approach to determining the intrinsic photon emission time is established within a general framework devoid of any arbitrary assumptions, with the parameters derived exclusively through data fitting. It is a fundamental mathematical principle that any analytical function can be represented by a Taylor expansion within a certain range. Consequently, we can express the general relationship between the intrinsic photon emission time 
$\Delta t_{\rm{in}}$ and energy $E$ as a Taylor expansion involving terms up to $E^n$:
\begin{equation}
\label{generalDelta}
\Delta t_{\rm{in}}=\Delta t_{\rm{in},c}+\alpha E+\beta E^2+\gamma E^3+\cdots,    
\end{equation}
where $\Delta t_{\rm{in},c}=\mu$ represents the constant term. By incorporating additional $E^2$ and $E^3$ terms in the intrinsic photon emission time, 
$\Delta t_{\rm{in}}$ can be regarded as a Taylor expansion of an analytical function within a certain level of precision. Consequently, our model integrating both the Lorentz violation term and the energy dependence of intrinsic emission time can be viewed as a specific instance of this general framework, eliminating the necessity for arbitrary assumptions regarding the parameters (i.e., the coefficients) and deriving them solely through data fitting.
Moreover, we demonstrate that by applying the aforementioned general expression of $\Delta t_{\rm{in}}$ up to the $E^3$ term to fit the data, it is notable that the coefficients $\beta$ and $\gamma$ converge toward 0 through data fitting of 17 events, as depicted in Fig.~\ref{energy_taylor}. Consequently, our model of intrinsic emission time can be considered a rational framework capable of accommodating all GRB photon data in Table~\ref{HighEnergyPhotons} for the analysis of Lorentz violation by incorporating the energy dependence of intrinsic emission time.

\begin{figure}[htbp] 
    \centering
	\includegraphics[width=0.65\textwidth]{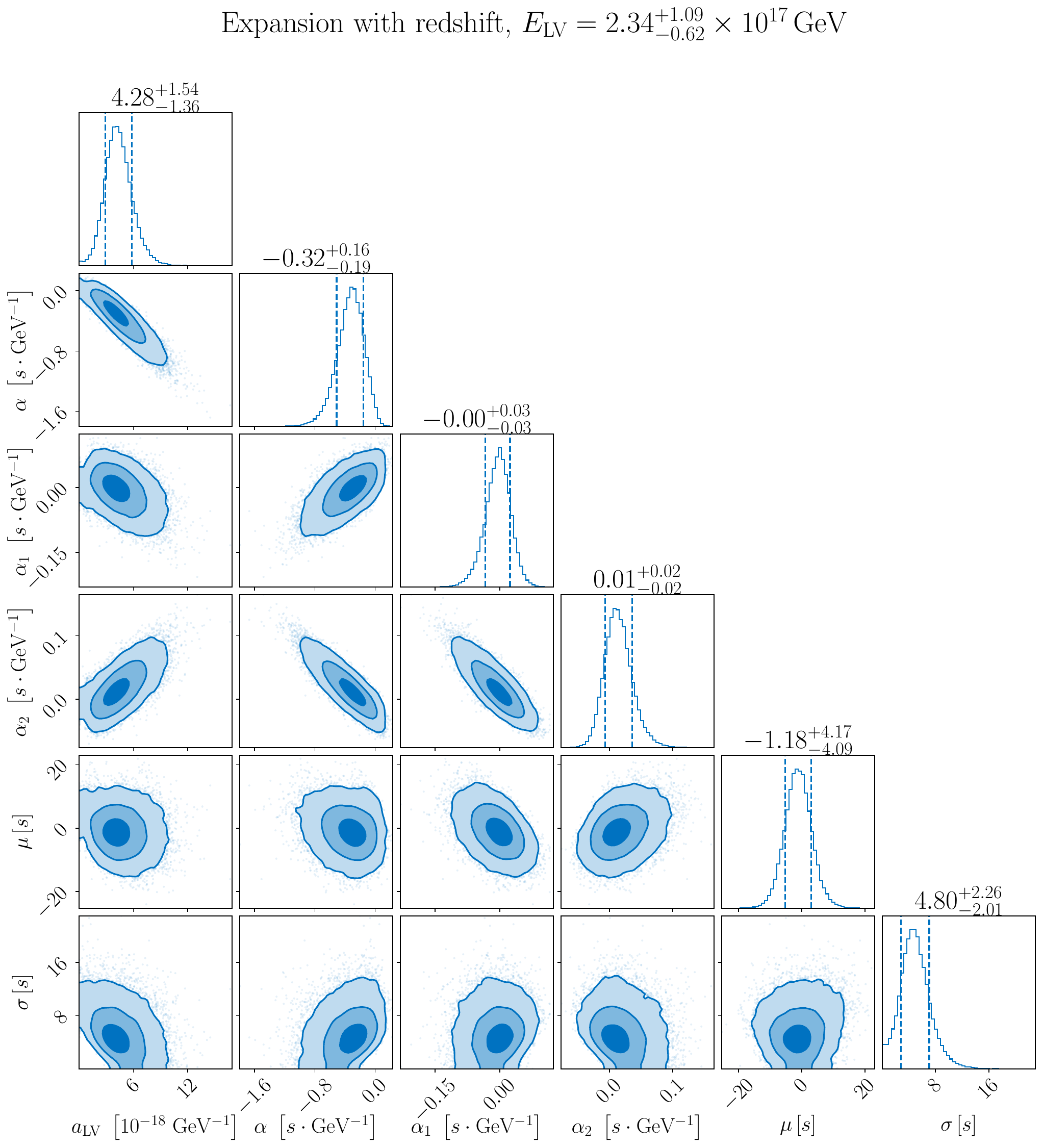} 
	\caption{
\reply{The analyzing results for the general expression of the intrinsic time-delay model with respect to redshift, as given in Eq.~\ref{refined_model}. The priors for $a_{\rm LV}$, $\alpha$, $\mu$ and $\sigma$ are the same as in Eq.~\ref{model}, and $\alpha_1$ and $\alpha_2$ are sampled from the uniform distribution $U \left[-30, 30 \right] \ {\rm s \cdot GeV^{-1}}$.}} 
	\label{redshift_expand} 
\end{figure}

\reply{
On the other hand, GRB engines may depend on the redshift $z$, which could also influence the intrinsic emission mechanism. Therefore,
we also perform the expansion of intrinsic time-delay model with respect to the redshift $z$, 
\begin{equation}
    \begin{aligned}
    \label{refined_model}
    \Delta t_{\mathrm{in}} &= \alpha(z) E_{\rm high,s} + \Delta t_{\rm in, c}, \\
    \alpha(z) &= \alpha + \alpha_1 (z - \Bar{z}) + \alpha_2 (z - \Bar{z})^2 + \cdots ,
\end{aligned}
\end{equation}
where $\alpha(z)$ denotes the expansion series on $z$, and $\Bar{z}$ refers to the averaged redshifts for all analyzed high-energy photons. 
Higher-order terms in $\alpha(z)$ account for the redshift dependence of the GRB's intrinsic properties. The fitting results are shown in Fig.~\ref{redshift_expand}. The estimated coefficients for the two higher order terms (i.e. $\alpha_1$ and $\alpha_2$) are nearly 0, while estimates for other parameters align with the results in Fig.~3.  The uncertainties for the previous four parameters $a_{\rm LV}$, $\alpha$, $\mu$, and $\sigma$ are increased.  Meanwhile, the LV energy scale $E_{\rm LV} = 2.34^{+1.09}_{-0.62} \times 10^{17}$ GeV is obtained, which is also consistent with the results in Fig.~3. Considering the limited data available at current stage, the results indicate that the leading-order expansion of redshift is sufficient for intrinsic time-delay model. The effects of higher-order terms still need to be explored with more GRB data in future. 
}

Based on the numerical results obtained in this study, we find that while the parameter \(\alpha\) is nonzero, the value of \(\mu\) is consistent with 0 within the margins of error. This indicates that \(\mu\) is relatively small, suggesting a potentially negligible effect in the context of our analysis. However, it is important to note that the energy dependence of intrinsic emission times for high-energy photons cannot be overlooked. This dependence may have significant implications for our findings and warrants further investigation.

We also find that when we adopt our full model, incorporating all four parameters \(a_{\mathrm{LV}}\), \(\alpha\), \(\mu\), and \(\sigma\) to be fitted from the data, we achieve consistent results for both the 14-photon dataset and the 17-photon dataset. However, when we fix \(a_{\mathrm{LV}} = 0\) (which corresponds to the three-parameter case), thereby assuming that Lorentz invariance is not violated, we obtain the main physical parameter $\alpha \sim 0.13 ~( \rm{s} \cdot {\rm GeV}^{-1})$  for the 14-photon dataset and $\alpha \sim 0.03 ~( \rm{s} \cdot {\rm GeV}^{-1})$ for the 17-photon dataset. This discrepancy indicates an inconsistency, highlighting the challenges in reconciling the three remarkable photons with the earlier 14-photon dataset under the assumption of Lorentz invariance.

In addition to the time-delay analysis of Lorentz violation from GRB photons, there are also potential signals arising from threshold anomalies caused by Lorentz violation in the interactions of high-energy GRB photons with the extragalactic background light (EBL). 
Features related to Lorentz violation with the remarkable LHAASO multi-TeV photon have been discussed in the literature~~\cite{Li:2022wxc,Li:2023rgc,Li:2023rhj},with a suggestion~\cite{Li:2023rhj} that \(E_{\mathrm{LV}} < 0.1 E_{\mathrm{P}}\) remains a viable explanation for the LHAASO results concerning the EBL interactions~\cite{LHAASO:2023lkv}. This perspective is consistent with the conclusions drawn in the present work based on the time-delay analysis.

The factors influencing gamma-ray absorption due to interactions with extragalactic background light (EBL) are indeed complex. For a reliable analysis, it is crucial to obtain detailed information about the EBL as well as the explicit energy spectrum of photons emitted from the GRB source. Variations in the observed high-energy photon spectrum can frequently be attributed to the intrinsic hardness or softness of the photon spectrum in the GRB source frame. Consequently, significant ambiguities arise when attempting to extract the Lorentz violation scale based on the absorption of GRB photons by the EBL.
From a theoretical perspective, the predicted signatures of Lorentz violation can vary substantially across different theoretical frameworks. For instance, within the context of doubly special relativity \cite{Amelino-Camelia:2002cqb, Amelino-Camelia:2002uql, Amelino-Camelia:2000stu}, one may observe threshold shifts in particle interactions without the presence of threshold anomalies. This allows for the possibility of using Lorentz violation to explain the observations of 10-TeV scale photons through these threshold shifts. As a result, the Lorentz violation features associated with EBL interactions may primarily serve to constrain specific Lorentz violation models, rather than providing a comprehensive assessment of the overall effects of Lorentz violation.

In this paper, we focus on the analysis of three ``remarkable GRB photons" in addition to previously examined cases, with the goal of establishing a clear and coherent framework for understanding high-energy events within the context of our proposed approach. By concentrating on these specific instances, we aim to demonstrate how our methodology offers a self-consistent interpretation of new observations. To further validate our findings and deepen our understanding of the implications of Lorentz invariance violation, additional investigations will be essential. We are actively progressing with the analysis of additional datasets from Fermi-LAT and LHAASO, and we anticipate that the outcomes will reinforce the conclusions presented here. This is because the results are derived using the same analytical methods, enabling us to capture a broader range of emission dynamics and thereby enhance our sensitivity to potential signals of Lorentz invariance violation.

\section{Summary}
In this study, we have enhanced the intrinsic time-delay model by incorporating the energy dependence of emission time \cite{plb138951}. Initially, we applied this model to the GRB data utilized in a previous study \citep{Xu:2016zxi, Xu:2016zsa} and achieved consistent results. Subsequently, we integrated the newly observed high-energy photons from the GeV to TeV bands by FGST, MAGIC, and LHAASO with the 14 high-energy photons observed by FGST in label I. Across all scenarios, we consistently observed a physical scenario where high-energy photons are emitted earlier than low-energy photons, with the LV energy scale estimated to be approximately $E_{\rm LV} \sim 3 \times 10^{17}$~GeV. 
Finally, our analysis of all 17 high-energy photons ranging from 33.58 GeV to 14.04 TeV at the source frame of 10 GRBs observed by FGST, MAGIC, and LHAASO reaffirmed the same physical scenario, indicating an LV energy scale of 
$E_{\rm LV} = 3.00^{+ 1.11}_{-0.64} \times 10^{17}$~GeV. 
After further taking into account higher-order terms of redshift dependence in the intrinsic emission time, we still obtain the Lorentz violation energy scale $E_{\rm LV} = 2.34^{+1.09}_{-0.62} \times 10^{17}$ GeV, which corresponds to   $1/E_{\rm LV} = 0.42^{+0.15}_{-0.14} \times 10^{-17}$ GeV$^{-1}$ as depicted in Fig.~6. Accordingly, the null hypothesis of dispersion-free vacuum $E=pc$ (or, equivalently, the constant light speed $v_{\gamma}=c$) is rejected at a significance level of 3.1$\sigma$ or higher.

\afterpage{
  \clearpage
  \thispagestyle{empty}
  \begin{sidewaystable}
  \vspace{0cm}
    \centering
\caption{
High-energy photon ($>$10 GeV) data from FGST, MAGIC, and LHAASO. Label I refers to the data from \citep{Xu:2016zsa, Xu:2016zxi}, while label II represents the new data from FGST \citep{Lesage:2023vvj}, MAGIC \citep{MAGIC:2019lau, MAGIC:2020egb}, and LHAASO \citep{LHAASO:2023lkv}. For the GRBs in Label I, 
the redshift $z$, 
the energy value in the observer frame $E_{\rm high,o}$, 
and the arrival time for high-energy photons
$T_{\rm high,o}$ can be obtained from previous studies \citep{Greiner:2009pm, McBreen:2010ni, 2009GCN98731C, DElia:2010fwa, 2010GCN106061C, 2013GCN144551L,Atwood:2013dra, Zhang:2014wpb,  Ruffini:2014fba, 2016GCN19411R, 2016GCN194191T, fermi_data_access}. For the GRBs in label II, we include GRB 221009A from FGST \citep{Lesage:2023vvj, Zhu:2022usw}, GRB 190114C from MAGIC \citep{MAGIC:2019lau, MAGIC:2020egb}, and GRB 221009A from LHAASO \citep{LHAASO:2023lkv}. Specifically for LHAASO data, we utilize the reconstructed energy as depicted in Fig. 1 of the respective publication. The arrival time of low-energy photons 
$T_{\rm low,o}$ is determined using the same criteria as in previous studies \citep{Xu:2016zxi, Xu:2016zsa}, which involves identifying the first main significant peak of the GBM data in the 8-260 keV band. For the GRBs in label I, we retain the same values as reported in previous studies \citep{Xu:2016zxi, Xu:2016zsa,Liu:2018qrg}. The values of 
$T_{\rm low,o}$ 
for the GRBs in label II are presented in Fig.~\ref{GBM}. The trigger time for each GRB is set as time 0. The values of 
$\Delta t_{\rm obs}/({1+z})$, the energy value of high-energy photons in the source frame 
$E_{\rm high,s}$, 
and 
$K_1$ 
are obtained with associated uncertainties as discussed in the main text.
}
  \label{HighEnergyPhotons}
    \begin{tabular}{ccllllllll}
    \toprule
    Label & Observatory & GRB   & $z$     & $E_{\rm high,o}$ (GeV)   & $T_{\rm high,o}$ (s) & $T_{\rm low,o}$ (s) & $\frac{\Delta t_{\rm obs}}{1+z}$ (s) & $E_{\rm high,s}$ (GeV)  & $K_1$ ($10^{18} \ {\rm s} \cdot {\rm GeV}$) \\
    \midrule
    I & FGST  & GRB 080916C & 4.35  & 12.4 $\pm \ 1.24$  & 16.55 & 5.98 $\pm \ 5.00$ & 1.98 $\pm \ 0.93$ & 66.34 $\pm \ 6.63$ & 4.45 $\pm \ 0.45$ \\
    I & FGST  & GRB 080916C & 4.35  & 27.4 $\pm \ 2.74$ & 40.51 & 5.98 $\pm \ 5.00$ & 6.45 $\pm \ 0.93$ & 146.60 $\pm \ 14.66$ & 9.83 $\pm \ 0.99$ \\
    I & FGST  & GRB 090510 & 0.903 & 29.9 $\pm \ 2.99$ & 0.83  & -0.03 $\pm \ 5.00$ & 0.45 $\pm \ 2.63$  & 56.90 $\pm \ 5.69$ & 7.19 $\pm \ 0.72$ \\
    I & FGST  & GRB 090902B & 1.822 & 39.9 $\pm \ 3.99$ & 81.75 & 9.77 $\pm \ 5.00$ & 25.5 $\pm \ 1.77$ & 112.60 $\pm \ 11.26$ & 12.84 $\pm \ 1.29$ \\
    I & FGST  & GRB 090902B & 1.822 & 11.9 $\pm \ 1.19$ & 11.67 & 9.77 $\pm \ 5.00$ & 0.67 $\pm \ 1.77$ & 33.58 $\pm \ 3.36$ & 3.83 $\pm \ 0.39$ \\
    I & FGST  & GRB 090902B & 1.822 & 14.2 $\pm \ 1.42$ & 14.17 & 9.77 $\pm \ 5.00$ & 1.56 $\pm \ 1.77$ & 40.07 $\pm \ 4.01$ & 4.57 $\pm \ 0.46$\\
    I & FGST  & GRB 090902B & 1.822 & 18.1 $\pm \ 1.81$ & 26.17 & 9.77 $\pm \ 5.00$ & 5.81 $\pm \ 1.77$ & 51.08 $\pm \ 5.11$ & 5.83 $\pm \ 0.59$ \\
    I & FGST  & GRB 090902B & 1.822 & 12.7 $\pm \ 1.27$ & 42.37 & 9.77 $\pm \ 5.00$ & 11.55 $\pm \ 1.77$ & 35.84 $\pm \ 3.58$ & 4.09 $\pm \ 0.41$ \\
    I & FGST  & GRB 090902B & 1.822 & 15.4 $\pm \ 1.54$ & 45.61 & 9.77 $\pm \ 5.00$ & 12.70 $\pm \ 1.77$ & 43.46 $\pm \ 4.35$ & 4.96 $\pm \ 0.50$ \\
    I & FGST  & GRB 090926A & 2.1071 & 19.5 $\pm \ 1.95$ & 24.84 & 4.32 $\pm \ 5.00$ & 6.60 $\pm \ 1.61$ & 60.59 $\pm \ 6.06$ & 6.51 $\pm \ 0.66$ \\
    I & FGST  & GRB 100414A & 1.368 & 29.7 $\pm \ 2.97$ & 33.37 & 0.29 $\pm \ 5.00$ & 13.97 $\pm \ 2.11$ & 70.33 $\pm \ 7.03$ & 8.68 $\pm \ 0.87$\\
    I & FGST  & GRB 130427A & 0.3399  & 72.6 $\pm \ 7.26$ & 18.64 & 0.54 $\pm \ 5.00$ & 13.51 $\pm \ 3.73$ & 97.28 $\pm \ 9.73$ & 8.99 $\pm \ 0.90$\\
    I & FGST  & GRB 140619B & 2.67  & 22.7 $\pm \ 2.27$ & 0.61  & 0.10 $\pm \ 5.00$ & 0.14 $\pm \ 1.36$ & 83.31 $\pm \ 8.33$ & 7.92 $\pm \ 0.80$\\
    I & FGST  & GRB 160509A & 1.17  & 51.9 $\pm \ 5.19$ & 76.51 & 13.92 $\pm \ 5.00$ & 28.84 $\pm \ 2.30$ & 112.62 $\pm \ 11.26$ & 14.20 $\pm \ 1.43$\\
    II & FGST  & GRB 221009A & 0.151 & 99.3 $\pm \ 9.93$ & 241.30 & 251.33 $\pm \ 5.00$  & -8.71 $\pm \ 4.35$ & 114.29 $\pm \ 11.43$ & 6.15 $\pm \ 0.62$\\
    II & MAGIC & GRB 190114C & 0.4245 & 1070 $\pm \ 160.5 $ & 73.60  & 2.59 $\pm \ 5.00$  & 48.86 $\pm \ 3.51$ & 1524.21 $\pm \ 228.63$  & 157.30 $\pm \ 23.65$\\
    II & LHAASO & GRB 221009A & 0.151 & 12200 $\pm \ 2440$ & 643.00 & 251.33 $\pm \ 5.00$  & 340.19 $\pm \ 4.35$ & 14042.20 $\pm \ 2808.44$ & 755.66 $\pm \ 151.31$\\
    \bottomrule
    \end{tabular}
  \end{sidewaystable}
  \clearpage
}


\begin{acknowledgments}
The authors thank Zhenwei Lyu, Jie Zhu, Hao Li, and the anonymous referee for helpful comments. This work is supported by National Natural Science Foundation of China under Grants No.~12335006 and No.~12075003. This work is also supported by High-performance Computing Platform of Peking University.
\end{acknowledgments}

\software{bilby \citep{Ashton:2018jfp, Romero-Shaw:2020owr}}

\clearpage
\bibliographystyle{aasjournal}

\bibliography{magic.bib}{}
\end{document}